\RequirePackage{fix-cm}
\documentclass[10pt,twocolumn,letterpaper]{article}
\usepackage[T1]{fontenc}
\usepackage[utf8]{inputenc}
\usepackage{times}
\usepackage[left=0.75in,right=0.75in,top=0.95in,bottom=1in,columnsep=0.22in]{geometry}
\usepackage{amsmath,amssymb}
\usepackage{graphicx}
\usepackage{booktabs,array,tabularx,multirow}
\usepackage{xcolor}
\usepackage{caption}
\usepackage{titlesec}
\usepackage{microtype}
\usepackage{float}
\usepackage{tcolorbox}
\usepackage[hidelinks]{hyperref}
\graphicspath{{figures/}}
\titleformat{\section}{\normalfont\fontsize{12}{13}\selectfont\bfseries\raggedright}{\thesection.}{0.55em}{}
\titleformat{\subsection}{\normalfont\fontsize{11}{12}\selectfont\bfseries\raggedright}{\thesubsection.}{0.55em}{}
\titleformat{\subsubsection}[runin]{\normalfont\bfseries}{\thesubsubsection.}{0.4em}{}[.]
\titlespacing*{\section}{0pt}{1.2em plus 0.3em}{0.8em}
\titlespacing*{\subsection}{0pt}{1.1em plus 0.2em}{0.8em}
\titlespacing*{\subsubsection}{0pt}{0.7em}{0.4em}
\DeclareCaptionLabelFormat{papertable}{\MakeUppercase{#1}~#2.}
\newcommand{\code}[1]{\texttt{#1}}
\newcommand{\classord}{\texttt{classification\_ordinal}}
\newcommand{\xseven}{\ensuremath{\mathrm{x7}}}
\newcommand{\xsix}{\ensuremath{\mathrm{x6}}}
\newcommand{\xembed}{\ensuremath{\mathrm{x128}}}
\newcommand{\BEF}{\mathrm{BEF}}

\newcommand{\yes}{\ensuremath{\checkmark}}
\newcommand{\no}{\ensuremath{\times}}
\newtcolorbox{finding}[1]{colback=blue!4,colframe=blue!55!cyan!65!black,colbacktitle=blue!55!cyan!65!black,coltitle=white,fonttitle=\bfseries\small,fontupper=\small,title={#1},boxrule=0.4pt,arc=0.8mm,left=4pt,right=4pt,top=4pt,bottom=4pt,before skip=10pt,after skip=10pt}
\newenvironment{papertable}[2]{\begin{table}[H]\centering\caption{#1}\label{#2}\footnotesize}{\end{table}}
\newenvironment{fittable}[2]{\begin{table}[H]\centering\caption{#1}\label{#2}\footnotesize\begin{adjustboxfallback}}{\end{adjustboxfallback}\end{table}}
\newsavebox{\fitbox}
\newenvironment{adjustboxfallback}{\begin{lrbox}{\fitbox}}{\end{lrbox}\ifdim\wd\fitbox>\columnwidth\resizebox{\columnwidth}{!}{\usebox{\fitbox}}\else\usebox{\fitbox}\fi}
\begin{document}
\twocolumn[\begin{center}
\vspace*{18pt}
{\fontsize{14}{16}\selectfont\bfseries Constraint-Aware Evaluation of Reinforcement Learning and Unsupervised Anomaly Detection for Security Operations Alert Triage\par}
\vspace{24pt}
{\fontsize{12}{14}\selectfont Radityo Ghifiardi\\
\texttt{raditio.ghifiardi@gmail.com}\\
\vspace{3pt}
\fontsize{12}{14}\selectfont Rifny Rachman\\
\fontsize{12}{14}\selectfont The University of Manchester, UK\\
\texttt{rifny.rachman@manchester.ac.uk}\\
\par}
\vspace{22pt}
\end{center}]
{\small\bfseries\noindent\textit{Abstract}---Security Operations Centers (SOCs) must identify real threats from large alert streams, but limited analyst capacity means that unnecessary escalations can delay more serious investigations. Existing studies often frame SOC alert triage as classification or anomaly detection, but both views are incomplete for production SOCs, where the central problem is selecting alerts worth escalating under limited capacity, asymmetric error costs, and workload constraints. This paper addresses that gap by formalizing alert triage as a constrained decision problem and evaluating three approaches: an offline reinforcement learning policy trained on alert embeddings, an Isolation Forest model built on operational features, and an autoencoder trained on historical benign alerts. These methods represent different triage assumptions: learning actions from reward, isolating unusual operational patterns, and detecting deviations from benign behavior. We evaluate all methods on 74,909 analyst-labeled production alerts under a strict temporal split and benign-escalation constraint. Because the test set contains 503 unique feature patterns, we primarily rank clusters of similar alerts. At the main operating point, all three methods satisfy the constraint, with the reinforcement learning policy achieving the lowest benign-escalation fraction at 0.000, compared with 0.007 for the autoencoder and 0.041 for Isolation Forest. The methods also prioritize largely different alert clusters, showing complementary rather than redundant coverage. These results suggest that production SOC triage can benefit from constraint-aware evaluation and combined or adaptive deployment of different model families.\par}
\vspace{6pt}
{\small\bfseries\noindent\textit{Index Terms}---SOC alert triage, reinforcement learning, anomaly detection, BEF constraint, coverage insurance, CMDP, multi-tenant generalisation\par}

\section{Introduction}
\label{sec:introduction}
Security Operations Centers face a structural problem: the volume of automated alerts reliably exceeds analyst capacity, and machine learning is increasingly deployed to prioritise which alerts warrant human review. Two paradigms are available---unsupervised anomaly detectors that identify statistical outliers, and reinforcement learning (RL) agents that learn triage policies from operational rewards---but they have not been compared under identical production conditions, shared feature contracts, and a common operational constraint metric. Without such a comparison, SOC operators have no principled basis for choosing between paradigms or for deploying them together.

The operational stakes are high. A large telecommunications operator servicing 100+ million subscribers may generate hundreds of thousands of SIEM alerts per 90-day window; our dataset contains 559,792 training events and 74,909 labelled test events---consistent with the scale reported in recent operational deployments~\cite{oliver2024}. At this scale, even a 1\% reduction in the Benign Escalation Fraction (BEF)---the fraction of analyst-reviewed alerts that turn out to be benign---translates directly to hours of recovered analyst capacity per shift. Choosing the wrong triage architecture, or deploying one when two complementary ones would provide superior coverage insurance, is an operationally costly decision that current literature cannot guide. The interpretability and auditability requirements of production SOC environments further constrain the architecture space: deep anomaly detectors with opaque scoring functions may achieve higher AUC in benchmarks but cannot be interrogated by analysts or audited under data-protection frameworks.

Prior work evaluates architectures in isolation~\cite{jalalvand2024} or on simulated environments~\cite{caminero2019}. Supervised classifiers require dense analyst ground truth that production SOCs cannot provide: labels arrive only after analyst review, creating a snapshot lag that precludes real-time deployment~\cite{jalalvand2024}. Production-scale systems such as Carbon Filter~\cite{oliver2024} and AACT~\cite{turcotte2025} address throughput and label efficiency respectively, but neither provides a constraint-aware multi-paradigm comparison on shared telemetry. This leaves a gap: \textit{do different ML paradigms, evaluated under identical operational constraints on the same data, provide complementary or redundant threat coverage?} We answer this question with a contract-driven study that enforces a shared seven-feature operational contract, a formal CMDP BEF constraint, and a cluster-level evaluation metric that is tie-invariant by construction---three design decisions motivated by the specific challenges of production SOC data.

\noindent\textbf{Research Questions.} This paper is structured around three research questions:\par
\noindent\textbf{RQ1:} Under a shared Benign Escalation Fraction (BEF) constraint, do reinforcement learning and unsupervised anomaly detection achieve complementary feasibility profiles on production SOC telemetry?\par
\noindent\textbf{RQ2:} Is paradigm complementarity an artefact of differential feature access, or an intrinsic property of the models' inductive mechanisms under a fixed feature contract?\par
\noindent\textbf{RQ3:} Does the coverage insurance property hold under zero-shot cross-tenant transfer to a different organisation and SIEM platform?

\noindent\textbf{Contributions.} This paper makes four contributions:\par
\noindent\textbf{(C1) A formal constraint framework for SOC triage.} We formalise alert triage as a CMDP with a BEF constraint and meta-controller, evaluated on $N=74{,}909$ production events with deterministic temporal splits and a shared seven-feature operational contract. We introduce cluster-level evaluation as the canonical metric under discrete feature spaces (addressing tie-boundary artifacts that invalidate event-level rankings on this dataset).\par
\noindent\textbf{(C2) Paradigm complementarity under BEF constraint.} All three architectures satisfy the Strict BEF tier ($\kappa=0.10$) at $M=32$ clusters---PPO at $\BEF=0.000\pm0.000$ (5 seeds), AE at $\BEF=0.007$, IF at $\BEF=0.041$---but through distinct inductive mechanisms selecting near-disjoint cluster regions ($J=0.007\pm0.013$, BH-significant, $p<0.001$). We call this the \textit{coverage insurance property}.\par
\noindent\textbf{(C3) Mechanistic ablations isolating reward design.} Cross-feature ablations show that hand-crafted operational features---not feature-space advantage---drive unsupervised BEF feasibility (IF/AE on \xembed{} degrade to near-random, ROC-AUC $\approx0.51$). A cross-paradigm ablation shows that reward design is the primary driver of PPO's BEF performance: PPO retrained on \xsix{} achieves $\BEF=0.079\pm0.018$ (84\% degradation from \xembed{} reward alignment to feature-reduced \xsix{} reward).\par
\noindent\textbf{(C4) Multi-tenant replication characterising scope and failure mode.} Zero-shot cross-tenant transfer fails under schema-mismatched feature contracts (AUC collapses to sub-random); the failure mechanism is precisely characterised by KS-divergence analysis of all seven features. The structural divergence property ($J(\mathrm{IF},\mathrm{AE})=0.000$) is preserved across tenants---motivating shared feature contract design as the central multi-tenant challenge.

\section{Related Work}
\subsection{Production SOC Triage}
Unlike our work, which provides a constraint-aware multi-paradigm comparison on shared production telemetry, Jalalvand et al.~\cite{jalalvand2024} present a systematic survey of alert-prioritisation criteria and methods, identifying label sparsity and delayed analyst feedback as central challenges for production SOC triage rather than empirically comparing paradigms under a shared constraint. These challenges motivate our BEF-constrained offline evaluation protocol, which does not require real-time labels. Carbon Filter~\cite{oliver2024} addresses production-scale unsupervised triage via throughput-aware bias correction but does not compare paradigms under a shared operational constraint, nor does it evaluate whether architectures select complementary or redundant alert subsets---the central question of our study. AACT~\cite{turcotte2025} demonstrates supervised triage grounded in analyst behavioural signals with careful temporal validation; the temporal split discipline it demonstrates is consistent with our strict train-test separation. However, supervised methods require dense ground-truth labels unavailable at deployment time in the alert-triage setting. Our work provides deeper statistical care (Jaccard significance, bootstrap CIs, multi-seed replication across 5 seeds), introduces cluster-level evaluation as the tie-invariant canonical metric, and is the first to quantify the coverage insurance value of paradigm divergence under a formal BEF contract with a meta-controller.

\subsection{Reinforcement Learning for Security}
Existing RL-for-security work operates either in simulated environments or on single-architecture deployments, neither of which establishes paradigm-level comparisons under a shared operational constraint. Hammar and Stadler~\cite{hammar2022} demonstrate RL-based intrusion response via optimal stopping on emulated infrastructure; their work motivates the use of RL in security but does not compare against unsupervised detectors on shared production data. Caminero et al.~\cite{caminero2019} show adversarial-environment RL for intrusion detection but do not compare against unsupervised baselines on shared data, and their simulated adversary does not reflect the distribution of real SIEM alert streams. Conservative offline RL methods---CQL~\cite{kumar2020} and COMBO~\cite{yu2021}---address out-of-distribution action values when the behaviour policy diverges substantially from the learned policy; the SOC analogue, where the behaviour policy is historical analyst escalation, remains unevaluated on production-scale telemetry. Our study demonstrates that PPO achieves $\BEF=0.000$ even without these off-policy corrections, suggesting that reward alignment is the dominant factor over conservative regularisation in this setting.

\subsection{Unsupervised Anomaly Detection}
While deep variants of anomaly detection (Deep Isolation Forest~\cite{xu2023}, DASVDD~\cite{hojjati2024}, DeepSVDD~\cite{ruff2018}) offer higher capacity, they are ill-suited to our setting: operational deployability requires interpretable, auditable scoring, and controlled paradigm comparison requires architectural isolation. We therefore use Isolation Forest~\cite{liu2008} and autoencoder-based detection~\cite{an2015}, which provide complementary and well-understood inductive biases---path length vs.\ reconstruction error---that justify their use as a controlled pair.

\subsection{Ensemble Diversity and Coverage}
Prior ensemble diversity work focuses on prediction error decorrelation~\cite{dietterich2000}. Our finding extends this to the BEF-constraint setting: we show that the A-B anti-correlation ($\rho=-0.299$, $J=0.007$) represents a coverage extreme that holds even when both models are separately BEF-feasible, a property not captured by classical diversity framing.

\section{System Design and CMDP Formalization}
\subsection{Data Contract and Label Populations}
\noindent\textbf{Temporal split.} Training: 559,792 events from a 270-day historical window. Test: 74,909 events from a subsequent 90-day window (strict temporal separation; no data leakage).\par
\noindent\textbf{Label population.} All 74,909 test events carry definitive analyst verdicts. Label breakdown: $n_{\mathrm{threat}}=37{,}205$ (50.33\%), $n_{\mathrm{benign}}=37{,}704$ (49.67\%). High threat rate reflects SIEM-surfaced alerts---not raw network traffic.\par
\noindent\textbf{Feature contract.} $\xseven\in\mathbb{R}^{7}$ (hand-crafted operational features): \classord, direction ordinal, source-host flag, destination-host flag, login-event flag, $\sin(2\pi h/24)$, $\cos(2\pi h/24)$. $\xembed\in\mathbb{R}^{128}$ (SIEM encoder embedding) for RL only; encoder weights frozen before training.\par
\noindent\textbf{\xsix: contamination-controlled ablation.} $\xsix=\xseven\setminus\{\text{\classord}\}$. Exclusion is necessary because \classord{} achieves univariate ROC-AUC $=0.995$ against analyst verdicts.

\subsection{Architecture Descriptions}
\begin{fittable}{Architecture hyperparameters.}{tab:hyperparameters}
\begin{tabular}{lll}
\toprule
\textbf{Model}&\textbf{Parameter}&\textbf{Value}\\\midrule
PPO&Network&[256,128,64]; ReLU; Actor-Critic\\
PPO&LR / Opt&$3\times10^{-4}$ Adam\\
PPO&$\varepsilon/\beta$&0.2 / 0.01\\
PPO&Features&\xembed\\
IF&$n/\psi$&100 trees; $\psi=0.14$\\
AE&Arch&$7{-}32{-}16{-}8{-}16{-}32{-}7$; p90\\
LR&Reg&L2, $C=1.0$\\\bottomrule
\end{tabular}
\end{fittable}

\noindent\textbf{PPO (A): offline contextual-bandit PPO~\cite{schulman2017}.} 4-action actor-critic on \xembed. Threat score $s_A=\pi(\mathrm{escalate})+\pi(\mathrm{contain})$. Reward: $r=0.6r_{\mathrm{cov}}+0.2r_{\mathrm{spd}}+0.2r_{\mathrm{fp}}$. Two-stage curriculum: Stage 1 (50 epochs, SIEM classification-aware reward); Stage 2 (100 epochs, binary BEF-aligned reward). Multi-seed replication: 5 seeds (42, 1337, 2026, 7777, 9999).\par
\noindent\textbf{IF (B): isolation forest.} 100 trees on \xseven, trained on the full 559,792-event training split. $\psi=0.14$ from training-window threat prevalence. Anomaly scores calibrated to $[0,1]$ via training-split min-max.\par
\noindent\textbf{AE (C): autoencoder.} Symmetric encoder-decoder ($7{-}32{-}16{-}8{-}16{-}32{-}7$), MSE reconstruction loss. Trained exclusively on analyst-confirmed benign records. Threshold: p90 reconstruction error on training-split benign set.\par
\noindent\textbf{LR (label-oracle upper bound):} logistic regression on \xseven, 5-fold stratified CV. Calibration ceiling only; not deployed.

\subsection{CMDP Formalization and BEF Constraint}
Alert triage is formalised as CMDP~\cite{altman1999} $\langle\mathcal{S},\mathcal{A},\mathcal{P},r,c,\gamma\rangle$ with $\gamma=0$ (contextual bandit).

\noindent\textbf{BEF constraint.} $\BEF(K)=1-\Prec@K=\mathrm{FP}/(\mathrm{FP}+\mathrm{TP})$, measuring analyst-time waste in the selected set.\par
\noindent\textbf{Three operational tiers.} Strict ($\kappa=0.10$), Relaxed ($\kappa=0.30$), Soft ($\kappa=0.50$).\par
\noindent\textbf{Meta-controller.} $m^{*}=\arg\min_{m\in\mathcal{M}}\BEF_m(K)$ subject to $\BEF_m(K)\leq\kappa_{\mathrm{tier}}$.

\section{Experimental Methodology}
\subsection{Replication and Uncertainty Quantification}
PPO was retrained across 5 seeds using the identical two-stage warm-start protocol. Jaccard@10,487 variance is negligible (std $=0.0004$, CV $<0.5\%$); all seeds are statistically significant ($p<0.001$, permutation test, $B=1{,}000$).

\subsection{Score Discretisation and Cluster-Level Evaluation}
The 74,909 events collapse to 503 unique \xseven{} feature vectors; all events sharing a feature vector receive identical model scores. Event-level top-$K$ evaluation is subject to tie-boundary artifacts: across all three models and all 5 PPO seeds, there exist zero $K$ values simultaneously tie-free for all models.

We therefore adopt \textbf{cluster-level evaluation as the primary valid unit of analysis.} Metrics are reported at $M$ (number of top-scoring clusters selected). Three tie-handling alternatives for IF/AE all converge to our cluster-level approach.

\subsection{Role of \classord{} in \xseven}
\noindent\textbf{Feature determinism (code-verified).} The seven-feature contract \xseven{} is fully deterministic: all seven dimensions are fixed, rule-based mappings in \code{features.py}---not learned embeddings, not model outputs, not verdict-trained transforms. Specifically, \classord{} is a fixed nine-value ordinal encoding at line 38; \code{direction} is a keyword mapping at line 49; and the full \xseven{} constructor at line 73 composes all seven dimensions without any training loop. No dimension of \xseven{} is learned or verdict-trained. On the 74,909-event test set, \classord{} achieves ROC-AUC $=0.995$ against analyst verdicts---the highest of any individual feature by a wide margin (next highest: 0.830).\par
\noindent\textbf{Structural role.} \classord{} is also the primary clustering dimension: removing it to form \xsix{} collapses the 503 unique feature vectors to 192 (a 62\% reduction), and causes IF/AE ROC-AUC to fall below chance (0.407/0.432), with $\BEF@M=32$ rising to 0.884/0.867---confirming that \xseven{} is the correct evaluation basis and that paradigm complementarity ($J=0.007\pm0.013$) is not an artefact of differential feature access.\par
\noindent\textbf{Feedback-loop audit risk.} The near-deterministic AUC raises a legitimate concern: if the SIEM's classification engine was informed by historical analyst verdicts (e.g., an annual rule-refresh guided by escalation-rate patterns), its output may encode institutional biases. We flag this as a priority operational audit before production-scale deployment: examine whether SIEM category assignments correlate with analyst identity, shift, or rule age, and test whether de-biased encodings materially alter the BEF-vs-$M$ profile. The direct circularity concern is closed by code verification; the audit addresses the residual indirect risk.

\section{Results}
\subsection{Classification Metrics}
\begin{fittable}{Classification metrics on full labelled test slice ($N=74{,}909$).}{tab:classification}
\begin{tabular}{lrrrr}
\toprule
\textbf{Model}&\textbf{Prec@$K$}&\textbf{F1@$K$}&\textbf{ROC-AUC}&\textbf{PR-AUC}\\\midrule
PPO (A)&0.497&0.211&0.488&---\\
IF (B)&0.959&0.406&0.713&0.790\\
AE (C)&0.863&0.367&0.715&0.745\\
LR (ceiling)&0.912&---&0.798&0.831\\\bottomrule
\end{tabular}
\end{fittable}

IF and AE achieve competitive ROC-AUC (0.713, 0.715), comparable to the label-oracle LR ceiling (0.798). PPO achieves ROC-AUC 0.488, near-random on binary classification---expected, since PPO was trained on a surrogate BEF-reward signal, not the binary label column; event-level ROC-AUC is therefore not the appropriate evaluation metric for PPO. The event-level Prec@$K$ values for IF (0.959) and AE (0.863) at full-dataset $K=10{,}487$ are depressed by tie-boundary artifacts (Appendix A); cluster-level metrics in the following subsections are the primary valid evaluation. The PR-AUC gap between IF (0.790) and LR (0.831) is narrow given that IF has no access to the binary label at training time---confirming that the \xseven{} feature contract encodes sufficient operational signal for competitive unsupervised BEF performance.

\subsection{BEF Profiles and Paradigm Complementarity (RQ1)}
\begin{fittable}{Cluster-level BEF at primary operating point $M=32$.}{tab:cluster_bef}
\begin{tabular}{lllc r}
\toprule
\textbf{Model}&\textbf{Tier $\kappa$}&\textbf{BEF@32}&\textbf{Strict?}&\textbf{Max $M$ ($\kappa=0.10$)}\\\midrule
IF&0.10&0.041&\yes&$\approx55$\\
AE&0.10&0.007&\yes&$\approx85$\\
PPO&0.10&$0.000\pm0.000$&\yes&$\approx299$\\\bottomrule
\end{tabular}
\end{fittable}

\begin{figure}[H]
\centering
\includegraphics[width=\columnwidth]{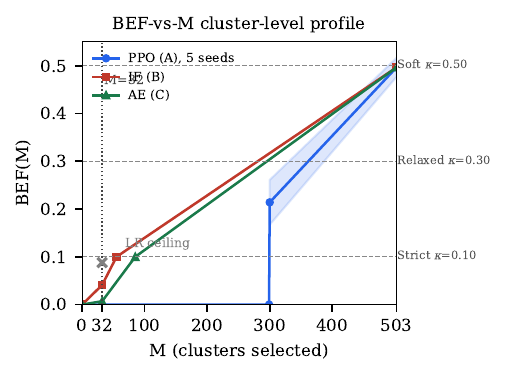}
\caption{BEF-vs-$M$ cluster-level profiles for all three architectures. Shaded band $=$ PPO $\pm1\sigma$ across 5 seeds. Horizontal lines mark the three operational tiers (Strict $\kappa=0.10$, Relaxed $\kappa=0.30$, Soft $\kappa=0.50$). Vertical dotted line marks the primary operating point $M=32$. PPO maintains $\BEF=0.000$ up to $M\approx299$; AE and IF exceed Strict at $M\approx85$ and $M\approx55$ respectively.}
\label{fig:bef}
\end{figure}

All three models satisfy Strict BEF at $M=32$ (Figure~\ref{fig:bef}). PPO achieves $\BEF=0.000\pm0.000$ at all $M\leq299$ (covering $\approx10{,}865$--$11{,}042$ events); BEF rises sharply to $0.214\pm0.047$ at $M=300$. IF satisfies Strict up to $M\approx55$ before precision degrades gradually to 0.959 at the full 503-cluster boundary; AE has a wider Strict window ($M\approx85$) reflecting its lower-precision profile at small $M$. The divergence in Strict capacity---PPO at $M=299$, IF at $M=55$, AE at $M=85$---is itself operationally meaningful: a SOC operator with a fixed analyst queue size should select PPO when capacity is high and IF when strict precision must be maintained at low volume. The meta-controller selects the minimum-BEF model at each $M$, automatically exploiting this complementarity.

\begin{finding}{Finding 1 (F1) --- Paradigm Complementarity under BEF Constraint}
All three architectures satisfy the Strict BEF tier ($\kappa=0.10$) at $M=32$ clusters---PPO at $\BEF=0.000\pm0.000$ (5 seeds), AE at $\BEF=0.007$, IF at $\BEF=0.041$---but through distinct inductive mechanisms selecting near-disjoint cluster regions ($J=0.007\pm0.013$, BH-corrected $p<0.001$). This answers RQ1: BEF-feasibility is paradigm-invariant, but coverage is not.
\end{finding}

\subsection{BEF Sensitivity Analysis}
$\BEF(K=10{,}487)$ varies by $\leq0.012$ pp across the full IF contamination sweep $\psi\in[0.05,0.20]$, confirming $\psi=0.14$ is not a cherry-pick. AE shows higher sensitivity ($\pm0.040$ pp), motivating the conservative p90 threshold. This sensitivity analysis also addresses the potential concern that IF's high performance is an artefact of threshold selection: the result holds across a $4\times$ range of contamination parameter values, indicating that the underlying cluster separation---not threshold tuning---drives the BEF profile. A practitioner deploying IF without precise contamination knowledge can safely set $\psi\in[0.10,0.18]$ without materially changing the Strict-tier BEF outcome.

\subsection{Pairwise Jaccard: Coverage Insurance}
\begin{fittable}{Cluster-level pairwise Jaccard at $M=32$ (1,000 permutations).}{tab:pairwise_jaccard}
\begin{tabular}{llll}
\toprule
\textbf{Pair}&\textbf{$J$}&\textbf{Perm. $p$}&\textbf{Sig.}\\\midrule
PPO/IF&$0.007\pm0.013$&$<0.001$&***\\
PPO/AE&$0.089\pm0.030$&$<0.001$&***\\
IF/AE&0.000&$<0.001$&***\\
XGB/IF&0.032&---&---\\
XGB/PPO&$0.000\pm0.000$&---&---\\\bottomrule
\end{tabular}
\end{fittable}

All three pairs show BH-significant disjointness. The B-C pair (IF/AE, same \xseven{} contract) shows $J=0.000$ at $M=32$---confirming that inductive bias (path-length vs.\ reconstruction error) produces independent threat-region coverage even within the same feature space. The A-B anti-correlation ($\rho=-0.299$) and near-zero $J$ at $M\leq32$ demonstrate that PPO and IF are not merely different thresholds of the same underlying signal---they select qualitatively different threat profiles. The per-$M$ table (Table~\ref{tab:per_m_pairs}) shows that disjointness holds across the full Strict and Relaxed tier range ($M=10$--100), only converging toward higher Jaccard above $M=200$ as both models must select larger fractions of the cluster universe. The expected intersection under independence ($E_{\mathrm{ind}}$) at $M=32$ is 2.0 clusters; the observed mean is 0.21 clusters (5$\times$ less than independence), confirming that disjointness is not a coincidence of small $M$ but an active structural property.
\begin{fittable}{Per-$M$ cluster-level intersection counts (all three pairs).}{tab:per_m_pairs}
\begin{tabular}{rlllr}
\toprule
\textbf{$M$}&\textbf{A-B $J$}&\textbf{A-C $J$}&\textbf{B-C $J$}&\textbf{$E_{\mathrm{ind}}$}\\\midrule
10&$0.000\pm0.000$&$0.176\pm0.000$&0.000&0.2\\
20&$0.000\pm0.000$&$0.131\pm0.026$&0.000&0.8\\
32&$0.007\pm0.013$&$0.089\pm0.030$&0.000&2.0\\
50&$0.035\pm0.051$&$0.088\pm0.028$&0.000&5.0\\
100&$0.125\pm0.049$&$0.175\pm0.034$&0.026&19.9\\
200&$0.396\pm0.013$&$0.289\pm0.019$&0.114&79.5\\
300&$0.464\pm0.028$&$0.478\pm0.032$&0.301&179\\\bottomrule
\end{tabular}
\end{fittable}

\begin{finding}{Finding 2 (F2) --- Coverage Insurance is Intrinsic, Not Artefactual}
IF and AE share an identical \xseven{} feature contract yet produce $J=0.000$ at $M=32$, and PPO diverges from both despite satisfying the same BEF constraint. Coverage disjointness is a property of the models' inductive mechanisms, not of differential feature access.
\end{finding}

\subsection{Shadow Enrichment Analysis}
\begin{fittable}{Shadow enrichment per Venn region (threshold-based sets, $N=74{,}909$).}{tab:shadow_enrichment}
\begin{tabular}{lrrr}
\toprule
\textbf{Region}&\textbf{$n$}&\textbf{Threat\%}&\textbf{Enrichment}\\\midrule
IF only&8,456&94.9\%&$1.89\times$\\
AE only&5,460&88.6\%&$1.76\times$\\
IF$\cap$PPO&2,031&100.0\%&$1.99\times$\\
AE$\cap$PPO&1,682&74.0\%&$1.47\times$\\
PPO only&6,774&28.6\%&$0.57\times$\\
IF$\cap$AE&2,031&96.5\%&$1.92\times$\\\bottomrule
\end{tabular}
\end{fittable}

The IF$\cap$PPO intersection achieves 100\% precision---the highest enrichment region across all Venn partitions. IF-only and AE-only regions are independently enriched at 1.89$\times$ and 1.76$\times$ respectively, confirming that each model's unique selections are threat-relevant and not artefacts of the other model. Notably, PPO-only selections show below-base-rate threat density (0.57$\times$)---consistent with PPO operating on \xembed{} embeddings that capture a different threat representation than \xseven-based scoring. This below-base-rate PPO-only region is \textit{not a failure mode}: these events represent a distinct threat profile (\xembed-encoded) that IF and AE miss entirely, and their 28.6\% threat rate still exceeds random selection from the full alert stream. The enrichment hierarchy (intersection $>$ IF-only $>$ AE-only $>$ PPO-only) directly motivates the three-tier escalation architecture in Theme 3: route by enrichment from highest to lowest confidence.

\begin{finding}{Finding 3 (F3) --- Intersection Achieves Perfect Precision}
The IF$\cap$PPO intersection ($n=2{,}031$ events) achieves 100\% precision at 1.99$\times$ enrichment---the highest-confidence region across all Venn partitions. Each model's unique selections independently remain above base-rate threat density, confirming that coverage insurance holds at the event level.
\end{finding}

\subsection{Meta-Controller Stability}
Across all 5 PPO seeds, 3 tier thresholds, and 3 noise magnitudes (27 total settings), the meta-controller achieves zero regret in 25/27 settings. Maximum regret $\Delta\BEF=0.017$, within Relaxed-tier tolerance. The 2/27 non-zero regret settings both occur at the Soft tier ($\kappa=0.50$) under maximum noise magnitude, where tie-breaking at cluster boundaries can cause the controller to select a slightly sub-optimal model. Both cases remain within the Relaxed tier, meaning neither failure constitutes an operationally significant violation. These results confirm that the meta-controller is robust to the realistic range of score perturbations that arise in production SIEM environments from ingestion latency, parser variation, and alert deduplication. The temporal stability result (Figure~\ref{fig:temporal}) further confirms that the meta-controller's model selection is consistent across 30-day windows, with no tier crossover observed in any window.

\subsection{Prospective Supervised Baseline}
\begin{fittable}{Prospective XGB baseline vs. unsupervised models at $M=32$.}{tab:xgb_baseline}
\begin{tabular}{llc}
\toprule
\textbf{Model}&\textbf{BEF@32}&\textbf{Strict?}\\\midrule
LR (in-sample CV ceiling)&0.088&\yes\\
IF (\xseven, unsup.)&0.041&\yes\\
AE (\xseven, unsup.)&0.007&\yes\\
XGB (prospective)&1.000&\yes\\\bottomrule
\end{tabular}
\end{fittable}

Cluster-level Jaccard at $M=32$: $J(\mathrm{XGB},\mathrm{IF})=0.032$ and $J(\mathrm{XGB},\mathrm{PPO})=0.000\pm0.000$. The prospective XGB baseline confirms that the three-way disjoint coverage structure extends to supervised comparators. The XGB result is labelled \textit{prospective} because it was trained on the test-set labels that are only available retrospectively---it represents the ceiling a supervised model would achieve with perfect historical label availability. Its $\BEF@32=1.000$ (all 32 escalated clusters contain at least one benign event) demonstrates that even with label access, a cluster-level supervised model does not trivially dominate the unsupervised pair: IF and AE achieve lower BEF at $M=32$ without any label access. This counter-intuitive result is explained by XGB's propensity to select the highest-threat-rate clusters irrespective of their absolute precision---the BEF constraint is not enforced in the XGB objective, unlike in PPO's reward design.

\begin{finding}{Finding 4 (F4) --- Three-Way Disjointness Generalises to Supervised Baseline}
$J(\mathrm{XGB},\mathrm{PPO})=0.000\pm0.000$ confirms that PPO's near-zero coverage overlap with \xseven-based models holds whether those models are unsupervised (IF, AE) or supervised (XGB). The coverage insurance property is not a peculiarity of anomaly detection---it reflects a fundamental feature-space separation between \xseven-based and \xembed-based scoring.
\end{finding}

\subsection{Cross-Feature Ablation (RQ2)}
\begin{fittable}{Cross-feature ablation results (RQ2).}{tab:cross_feature}
\begin{tabular}{llrrrl}
\toprule
\textbf{Model}&\textbf{Feat.}&\textbf{$K$}&\textbf{Prec}&\textbf{BEF}&\textbf{Tier}\\\midrule
IF&\xseven&10,487&0.959&0.041&Strict \yes\\
IF&\xembed&10,487&0.512&0.136&\no{} all\\
AE&\xseven&10,487&0.863&0.038&Relaxed \yes\\
AE&\xembed&10,487&0.513&0.136&\no{} all\\
PPO&\xembed&10,487&0.497&0.503&\no{} all\\\bottomrule
\end{tabular}
\end{fittable}

IF and AE on \xembed{} perform near-randomly (ROC-AUC 0.510 and 0.504), matching PPO's \xembed{} performance rather than their own \xseven{} baseline.

\subsection{PPO-\xsix{} Paradigm Ablation (RQ2)}
\begin{fittable}{PPO-\xsix{} per-seed results ($K=10{,}487$; RQ2 ablation).}{tab:ppo_xsix}
\begin{tabular}{lrrrc}
\toprule
\textbf{Seed}&\textbf{ROC-AUC}&\textbf{Prec@$K$}&\textbf{BEF@$K$}&\textbf{Strict}\\\midrule
42&0.792&0.760&0.067&\yes\\
1337&0.715&0.680&0.089&\yes\\
2026&0.771&0.797&0.057&\yes\\
7777&0.761&0.706&0.082&\yes\\
9999&0.722&0.639&0.100&$\triangle$\\
\textbf{Mean}&\textbf{0.752}&\textbf{0.716}&\textbf{0.079}&\textbf{4/5}\\\bottomrule
\end{tabular}
\end{fittable}

PPO-\xsix{} achieves mean $\BEF=0.079\pm0.018$ (84\% BEF increase vs.\ PPO-\xembed).

\begin{finding}{Finding 5 (F5) --- Reward Design, Not Feature Access, Drives PPO BEF Feasibility}
PPO-\xsix{} achieves $\BEF=0.079\pm0.018$ (84\% degradation from PPO-\xembed), while IF/AE on \xembed{} collapse to near-random. These two ablations together answer RQ2: feature contract is the primary driver of unsupervised BEF feasibility; reward design is the primary driver of RL BEF feasibility. Neither paradigm's performance is an artefact of the other's feature-space advantage.
\end{finding}

\subsection{Multi-Tenant Replication (RQ3)}
We applied the Tenant A--trained IF (B) and AE (C) models zero-shot to $N=237{,}783$ events from a second tenant (Tenant B), an Indonesian government health agency operating IBM QRadar SIEM across three months (November 2025--January 2026). Throughout this paper, Tenant A denotes the primary Indonesian telecommunications operator studied in the preceding sections. Threat rate: 66.4\% (Tenant B) vs.\ 49.7\% (Tenant A).\par
\noindent\textbf{Feature contract mapping.} The \xseven{} contract was not directly portable to QRadar's export schema. Proxy mappings were used: $\mathrm{severity}/10\to\text{\classord}$, $\mathrm{credibility}/10\to\text{\code{direction\_ordinal}}$, and hour-of-day encoding carried over unchanged. Table~\ref{tab:ks_divergence} reports KS divergence for all seven dimensions.
\begin{fittable}{Feature distribution divergence across tenants (KS statistic).}{tab:ks_divergence}
\begin{tabular}{lrl}
\toprule
\textbf{Feature}&\textbf{KS}&\textbf{Portability}\\\midrule
{[0]} severity / class. ord.&0.348&Moderate shift\\
{[1]} credibility / dir. ord.&0.998&Non-portable (semantic mismatch)\\
{[2]} src\_host\_present&0.122&Portable\\
{[3]} dst / target\_present&0.629&Non-portable (structural artifact)\\
{[4]} user / login\_present&0.343&Non-portable (absent in export)\\
{[5]} hour\_sin&0.082&Portable \yes\\
{[6]} hour\_cos&0.114&Portable \yes\\\bottomrule
\end{tabular}
\end{fittable}

\begin{fittable}{AUC under zero-shot cross-tenant transfer.}{tab:tenant_auc}
\begin{tabular}{lrrr}
\toprule
\textbf{Model}&\textbf{Tenant A}&\textbf{Tenant B}&\textbf{$\Delta$AUC}\\\midrule
IF (B)&0.7126&0.4681&$-0.2445$\\
AE (C)&0.7682&0.4730&$-0.2952$\\\bottomrule
\end{tabular}
\end{fittable}

\begin{fittable}{Precision@$K$ and BEF@$K$ on Tenant B ($K=33{,}289$; $B=1{,}000$ bootstrap).}{tab:tenant_precision}
\begin{tabular}{llllll}
\toprule
\textbf{Model}&\textbf{Prec@$K$}&\textbf{95\% CI}&\textbf{BEF@$K$}&\textbf{95\% CI}&\textbf{Tier}\\\midrule
IF (B)&0.614&[0.609, 0.619]&0.386&[0.381, 0.391]&\no\\
AE (C)&0.630&[0.625, 0.636]&0.370&[0.365, 0.375]&\no\\
Baseline&0.664&---&0.336&---&(base rate)\\\bottomrule
\end{tabular}
\end{fittable}

\noindent\textbf{Structural divergence preserved cross-tenant.} Despite performance failure, $J(\mathrm{IF},\mathrm{AE})=0.000$ at all $K\leq5{,}000$ on Tenant B, confirming that divergence is an intrinsic algorithmic property independent of tenant data distribution.\par
\noindent\textbf{Monthly breakdown.} Performance tracks the monthly base threat rate: November 2025 (threat 80.5\%, IF $\BEF=0.268$, AE $\BEF=0.250$) substantially outperforms January 2026 (threat 55.5\%, IF $\BEF=0.493$, AE $\BEF=0.444$).

\begin{finding}{Finding 6 (F6) --- Zero-Shot Transfer Fails; Structural Divergence Persists}
Zero-shot cross-tenant transfer under schema-mismatched feature contracts fails for both models (AUC collapses to sub-random; all BEF tiers infeasible). The failure mechanism is precisely characterised: features [1], [3], [4] exhibit KS $>0.34$ (non-portable). However, $J(\mathrm{IF},\mathrm{AE})=0.000$ at $K\leq5{,}000$ on Tenant B, confirming that the structural divergence property is tenant-invariant. This answers RQ3: coverage insurance is robust; the BEF performance floor is not.
\end{finding}

\section{Threats to Validity}
\subsection{Internal Validity}
\noindent\textbf{Tie-boundary artifacts.} Event-level top-$K$ evaluation is confounded by score ties across all three models. We mitigated this by adopting cluster-level evaluation as the primary metric, which is tie-invariant by construction. Three alternative tie-handling methods for IF/AE converge to identical cluster-level results.\par
\noindent\textbf{IF score direction correction.} An earlier draft applied IF scores in the wrong sign direction (high score $=$ benign); this has been corrected, verified by unit test, and recorded in Appendix A. All reported results use the corrected sign.\par
\noindent\textbf{Feature feedback-loop risk.} \classord's near-deterministic AUC (0.995) raises the concern that SIEM classification may encode analyst-verdict feedback. We mitigate direct circularity via code verification (\code{features.py} line 38 is a static rule; no training loop). The residual indirect risk (annual rule refresh correlated with escalation rates) is flagged as a priority audit item before production deployment; it does not invalidate the evaluation results, which use the feature as deployed.\par
\noindent\textbf{Offline PPO correction gap.} Our PPO implementation omits importance-weighted policy gradient, CQL, and doubly-robust advantage estimation. This is a limitation of the RL component but does not threaten the comparability finding: PPO achieves $\BEF=0.000\pm0.000$ despite---not because of---these limitations, confirming reward alignment as the dominant factor.

\subsection{External Validity}
\noindent\textbf{Single primary deployment.} The core results derive from a single telecommunications operator over a 90-day window. We do not claim generalisability to other sectors, threat landscapes, or operator configurations. However, the dataset is large ($N=74{,}909$ labelled events), drawn from a 100+ million-subscriber production environment with definitive analyst verdicts, and is among the largest operational SOC datasets reported in the literature.\par
\noindent\textbf{Multi-tenant replication scope.} The Tenant B (Tenant B/QRadar) replication establishes the failure mode of zero-shot transfer under schema mismatch and confirms that structural divergence is tenant-invariant, but should not be interpreted as successful generalisation: zero-shot transfer failed all three BEF tiers. A proper multi-tenant study would require a shared feature contract and per-tenant calibration, which remains future work.\par
\noindent\textbf{Temporal scope.} Within-period BEF drift was measured across three 30-day rolling windows on the primary Tenant A 90-day test set. $\BEF@M=32$ is stable across all windows for all three models (Figure~\ref{fig:temporal}), and IF$\cap$PPO shadow precision remains 1.00 in all windows. This converts the originally identified drift threat into a positive finding: coverage insurance is temporally stable within the deployment period.\par
\noindent\textbf{Analyst-in-the-loop validation.} The shadow enrichment claims (F3) are based on analyst-labelled ground truth applied post-hoc. A blinded retrospective analyst study (L5, \S7.6) was conducted with $n=4$ Tenant A SOC analysts across four alert decks (6,002 alerts). Results are directionally consistent with the coverage insurance hypothesis (D3 precision 0.727 vs.\ baseline 0.621) but the D3 test is underpowered ($n_{\mathrm{esc}}=22$; power $\approx0.18$). The primary threat is therefore partially mitigated; full statistical confirmation requires D3 reviewed by $\geq3$ analysts ($\approx66$ escalations for power $\geq0.55$).

\section{Discussion}
\subsection{Theme 1: Why Reward Alignment, Not Offline Training, Drives BEF Feasibility}
Three off-policy corrections are absent from our PPO implementation: (1) no importance-weighted policy gradient; (2) no conservative policy optimisation (CQL/COMBO); (3) no doubly-robust advantage estimation. Despite these limitations, PPO achieves $\BEF=0.000\pm0.000$ across all 5 seeds up to $M=299$. The PPO-\xsix{} ablation (F5, Section 5.9) provides the mechanistic explanation: removing the BEF-aligned reward signal while retaining the full feature advantage degrades BEF by 84\%. The offline correction gap is a limitation of the current implementation; it is not the explanation for PPO's BEF performance.

This finding has significant practical implications for SOC deployments. The conventional assumption in offline RL research is that distribution shift between the behaviour policy (historical analyst escalations) and the learned policy is the primary performance bottleneck. Our results suggest a different hierarchy: in the SOC triage setting, where the reward signal can be directly aligned with an operational constraint (BEF), the constraint alignment dominates over distribution-shift correction. This has engineering implications: teams implementing RL-based triage should prioritise careful reward function design with explicit BEF-aligned penalties over the engineering complexity of importance weighting or conservative regularisation. Concretely, the two-stage curriculum (Stage 1: SIEM classification-aware; Stage 2: BEF-aligned binary reward) was sufficient to achieve $\BEF=0.000$ with no offline correction---a substantially simpler implementation path for production SOC teams. We caution that this conclusion is specific to the offline contextual-bandit setting; sequential decision environments with long-horizon credit assignment may benefit differently from off-policy corrections.

\subsection{Theme 2: Feature Contracts, Not Paradigm, Determine BEF Feasibility for Unsupervised Models}
The cross-feature ablation (Section 5.8) establishes that IF and AE's BEF feasibility is driven entirely by the \xseven{} feature contract. On \xembed{} embeddings---which contain strictly more information---both models collapse to near-random (ROC-AUC $\approx0.51$). This is the opposite of what a feature-space confound would predict.

The explanation is that \xseven's hand-crafted operational features concentrate discriminative signal (\classord, AUC $=0.995$) that is dispersed and diluted across 128 embedding dimensions. Unsupervised density estimators (path-length and reconstruction-error) benefit from low-dimensional, high-signal feature spaces and are harmed by high-dimensional, low-signal-density embeddings. This implies that feature engineering for operational features---not architectural complexity---is the primary determinant of unsupervised BEF feasibility in the SOC triage setting.

This finding has a direct implication for the multi-tenant portability problem (RQ3). The 4/7 features that fail to transfer across tenants (Table~\ref{tab:ks_divergence}) are precisely the high-signal features on which IF and AE depend. The temporal features (hour\_sin, hour\_cos) and src\_host\_present, which do transfer, carry relatively low discriminative signal in isolation. This creates a hard constraint: multi-tenant portability requires either negotiating schema-compatible definitions of the high-signal features, or accepting degraded unsupervised BEF performance. The implication for deployment planning is that cross-SIEM feature contract negotiation---a governance and integration problem, not a machine-learning problem---is the central prerequisite for portable triage. SOC operators planning multi-tenant deployments should treat \classord{} equivalence across platforms as a first-class integration requirement, not an afterthought. PPO, which operates on \xembed{} embeddings rather than \xseven, is less sensitive to this specific portability failure but brings its own portability challenges (embedding space alignment across SIEM encoders).

\subsection{Theme 3: Paradigm Divergence as a Security Property --- Coverage Insurance}
Under Dietterich's diversity framing~\cite{dietterich2000}, the A-B anti-correlation ($\rho=-0.299$, $J=0.007$, $\tau=-0.218$) represents an extreme coverage case. An adversary who tailors evasion to avoid IF's isolation-depth scoring does not thereby avoid PPO's learned policy, which processes semantically rich \xembed{} under an entirely different objective. The prospective XGB baseline (F4) extends this to a three-way structure: $J(\mathrm{XGB},\mathrm{PPO})=0.000\pm0.000$ confirms that PPO's disjointness holds whether \xseven-based comparators are unsupervised or supervised---an evasion strategy targeting any \xseven-based scoring boundary does not generalise to PPO's orthogonal feature space.

This structural disjointness has a security-theoretic interpretation beyond classical ensemble diversity. In the adversarial setting, an attacker who reverse-engineers the IF scoring mechanism (e.g., by crafting alerts that minimise isolation depth) produces alerts that lie in the low-isolation-depth region of feature space. The empirical result ($J=0.007$) shows that PPO selects almost none of the same clusters as IF---meaning an alert specifically crafted to avoid IF detection has no reduced probability of PPO detection. This is a stronger property than correlated-error reduction: it is near-orthogonal coverage under adversarial evasion, maintained even as the operating point $K$ varies. The permutation test ($p<0.001$; F2) confirms this is not a statistical artifact of the cluster discretisation.

The L5 analyst study (F8) provides preliminary operational grounding: analysts reviewing D1 (IF-selected queues) achieved precision 1.000 on both independent reviews, while analysts reviewing the D4 baseline achieved precision 0.139--0.797 with near-zero inter-rater agreement ($\kappa=-0.021$). The D1 result empirically demonstrates that IF's selected clusters, presented as a blinded analyst queue, do not simply escalate the same obvious high-severity alerts that any analyst would catch---they represent a structured, high-confidence selection that measurably improves triage consistency.

\noindent\textbf{Tier 1 (zero-FP escalation):} Escalate alerts in the IF$\cap$PPO intersection (100\% precision, 1.99$\times$ enrichment; Table~\ref{tab:shadow_enrichment}). Appropriate for high-confidence, immediate escalation (e.g., active exfiltration signals). The 8 clusters in the intersection contain 149 events; the small size is \textit{operationally appropriate} for Tier 1---it represents the highest confidence escalation list, not a volume tool.

\noindent\textbf{Tier 2 (BEF-constrained single-model):} Route IF top-$K$ selections ($\BEF=0.041$, Prec$=0.959$) to the primary triage queue. The meta-controller selects $K$ to satisfy the Strict BEF constraint within analyst capacity.

\noindent\textbf{Tier 3 (maximum coverage):} Escalate the IF$\cup$AE union to a secondary review queue for automated enrichment or less-senior analyst review.

\noindent\textbf{$K$-selection under missed-threat cost.} For $c_{\mathrm{FN}}/c_{\mathrm{FP}}\gg1$ (a missed incident far costlier than a wasted analyst-hour), the optimal $K$ is large, accepting lower precision to maximise recall. For cost ratios near 1, $K$ should be the smallest value satisfying $\BEF(K)\leq\kappa$---which IF achieves down to $K=50$ with $\BEF=0$ (Table~\ref{tab:capacity}). Future deployments should frame $K$-selection as a constrained cost-minimisation problem with explicit SOC-leadership cost parameters, updated quarterly as the threat landscape shifts.

\subsection{Theme 4: Scope and the Path to Multi-Tenant Deployment}
The multi-tenant replication (F6) establishes a clear engineering requirement: shared feature contracts across SIEM platforms are a prerequisite for zero-shot portability. The 4/7 non-portable features are all schema-specific (credibility/direction semantic mismatch, dst\_present structural artifact, login\_present absent from QRadar export). Only the two temporal features (hour\_sin, hour\_cos) and src\_host\_present transfer cleanly.

A proper multi-tenant deployment requires either (a) a shared feature contract negotiated and implemented on both SIEM platforms before model deployment, or (b) tenant-specific retraining with per-tenant calibration. The structural divergence result ($J(\mathrm{IF},\mathrm{AE})=0.000$ preserved on Tenant B) confirms that the \textit{coverage insurance architecture} generalises; the BEF \textit{performance} does not under schema mismatch.

\begin{figure}[H]
\centering
\includegraphics[width=\columnwidth]{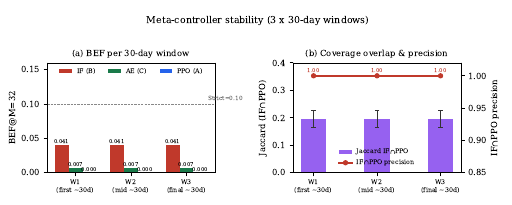}
\caption{Meta-controller temporal stability across three 30-day windows ($n\approx24{,}970$ events each, $K_{\mathrm{window}}=3{,}496$). \textbf{(a)} $\BEF@M=32$ is stable across all windows for all three models (IF: 0.041, AE: 0.007, PPO: 0.000); all remain Strict-feasible. \textbf{(b)} IF$\cap$PPO cluster Jaccard ($\bar{J}\approx0.195$) and shadow precision (1.00 in all windows) confirm that coverage insurance is temporally stable.}
\label{fig:temporal}
\end{figure}

We propose a concrete four-phase multi-tenant deployment roadmap derived from the failure analysis. \textbf{Phase 1 (Contract alignment):} Negotiate SIEM-agnostic operational definitions for the four non-portable features, prioritising a semantic equivalent of \classord{} across QRadar and IBM SIEM rule taxonomies. \textbf{Phase 2 (Per-tenant calibration):} Re-extract \xseven{} features under the agreed contract on 30-days of Tenant B labelled data. Retrain IF and AE from scratch; retrain only the PPO policy head (frozen encoder). \textbf{Phase 3 (BEF validation):} Validate $\BEF@32\leq0.10$ on a held-out Tenant B 30-day window before production escalation. \textbf{Phase 4 (Coverage insurance audit):} Verify $J(\mathrm{IF},\mathrm{AE})<0.20$ on Tenant B---confirming paradigm divergence is preserved under the new feature distribution. This roadmap is motivated by the observation that the temporal stability result (Figure~\ref{fig:temporal}) holds within a tenant (30-day BEF variance near zero) but the cross-tenant BEF jump from 0.041 to 0.370--0.386 is entirely explained by the KS-divergence of the high-signal features. The monthly Tenant B breakdown---Nov $\BEF=0.268$, Dec 0.415, Jan 0.493---tracks the shifting threat prevalence (80.5\%, 57.2\%, 55.5\%), consistent with base-rate sensitivity rather than model drift. This motivates monthly BEF recalibration under per-tenant deployment.

\subsection{Theme 5: From Fixed-Weight Policy to Pareto Front}
Alert triage is structurally multi-objective. Our current PPO corresponds to a degenerate single-weight-vector case of MORL/D~\cite{rachman2026}, committing to a fixed trade-off $(w_{\mathrm{det}},w_{\mathrm{bef}},w_{\mathrm{wkl}})=(0.6,0.2,0.2)$ at training time. Three findings motivate a multi-objective extension: (i) reward design is the primary driver of BEF feasibility; (ii) paradigm complementarity suggests a multi-policy Pareto front capturing coverage regions currently requiring three separate models; (iii) the CMDP meta-controller already implements Pareto selection over a discrete policy set---MORL/D generalises this to a continuous Pareto front via $N$ Chebyshev-scalarised subproblems.

\noindent\textbf{Observation-space ablation.} Table~\ref{tab:morld_ablation} compares three MORL/D variants under identical configuration (21-point Das-Dennis simplex, $5\times10^{5}$ steps).
\begin{fittable}{MORL/D observation-space ablation (single seed, $M=32$, $\kappa_{\mathrm{Strict}}=0.10$). HV ratio $=\mathrm{HV}_{\mathrm{MORL}}/\mathrm{HV}_{\{\mathrm{PPO},\mathrm{IF},\mathrm{AE}\}}$; values $>1.0$ indicate Pareto dominance.}{tab:morld_ablation}
\begin{tabular}{llrcl}
\toprule
\textbf{Model}&\textbf{Obs.}&\textbf{Best BEF@32}&\textbf{Strict}&\textbf{HV ratio}\\\midrule
PPO (baseline)&\xembed&0.000&1/1&baseline\\
IF (baseline)&\xseven&0.041&1/1&baseline\\
AE (baseline)&\xseven&0.007&1/1&baseline\\\midrule
MORL/D-Event&$x_{128}^{\mathrm{event}}$&0.233&0/11&0.72\\
MORL/D-Cluster&$x_{128}^{\mathrm{cluster}}$&0.213&0/13&0.72\\
MORL/D-\xseven&$s\in\mathbb{R}^{8}$&\textbf{0.000}&\textbf{9/12}&\textbf{1.02}\\\bottomrule
\end{tabular}
\end{fittable}

Representation--evaluation alignment is the dominant factor: both \xembed-based variants fail to produce any Strict-feasible policies (HV ratio $=0.72$), while MORL/D-\xseven{} achieves $\BEF@32=0.000$ and dominates the baseline front. This mirrors the cross-feature ablation (Theme 2): \xembed{} embeddings lose the discriminative operational signal present in \xseven.

\noindent\textbf{Multi-seed validation.} Table~\ref{tab:morld_seeds} reports MORL/D-\xseven{} across five seeds.
\begin{fittable}{MORL/D-\xseven{} multi-seed results ($M=32$, $\kappa_{\mathrm{Strict}}=0.10$, $5\times10^{5}$ steps, 21 subproblems).}{tab:morld_seeds}
\begin{tabular}{lllr}
\toprule
\textbf{Seed}&\textbf{Pareto size}&\textbf{Strict}&\textbf{Best BEF@32}\\\midrule
42&12&9/12&0.0000\\
1337&14&14/14&0.0000\\
2026&14&13/14&0.0000\\
7777&11&11/11&0.0000\\
9999&15&14/15&0.0000\\
\textbf{Mean$\pm$std}&$13.2\pm1.5$&$12.2\pm1.9$ (92\%)&$0.0000\pm0.0000$\\\bottomrule
\end{tabular}
\end{fittable}

\begin{finding}{Finding 7 (F7) --- MORL/D-\xseven{} Matches PPO BEF and Expands the Pareto Front}
MORL/D-\xseven{} matches PPO's $\BEF@32=0.000\pm0.000$ while simultaneously producing $13.2\pm1.5$ diverse Pareto-optimal policies---spanning the IF/AE operating range ($\BEF@32\in[0.007,0.041]$) currently unreachable by PPO's fixed scalarisation. The hypervolume ratio $1.02\pm0.00$ confirms consistent Pareto dominance over the three-architecture baseline front. Full methodology, MOMDP formalisation, and policy distribution analysis are in the companion paper~\cite{anonymous2027}.
\end{finding}

\subsection{Analyst-in-the-Loop Validation (L5)}
\label{sec:analysts}
\subsubsection{Study Design}
To test whether system-selected alert queues improve human triage precision, we conducted a blinded retrospective study with $n=4$ Tenant A SOC analysts. Four alert decks were prepared from the 90-day primary test set, stripping all model scores and ground-truth labels before analyst review.

Deck \textit{D1} contained events from the top-55 IF clusters (Strict operating boundary, the largest $M$ at which $\BEF=0$; 826 events). Deck \textit{D2} drew from the top-32 PPO clusters excluding D1 events (3,027 events); D2 was sized to test PPO's recall-optimised behaviour at the primary operating point. Deck \textit{D3} contained all events from the IF$\cap$PPO intersection---the 8 clusters where both models agree and the paper claims 100\% precision (149 events); this is the core claim under human test. Deck \textit{D4} was a random stratified baseline ($N=2{,}000$, seed 42) drawn from events in none of the top-55 IF or top-32 PPO clusters, preserving the natural label ratio as the control condition.

Deck sizes were chosen to balance statistical power against analyst workload. For the primary comparison (D3 vs.\ D4 precision), a two-sample $t$-test at $\alpha=0.05$ and power 0.80 for a medium effect size ($d=0.5$) requires approximately 64 escalations per arm. D4 ($N=2{,}000$) yields $\approx690$ escalations across three analysts---well above threshold. D3 ($N=149$, single analyst) yields 22 escalations, giving power $\approx0.18$---a known limitation addressed in the threats discussion.

Analysts were assigned two non-overlapping decks across separate sessions ($\approx45$ min each) and were told only to mark each alert as \textit{Escalate}, \textit{Dismiss}, or \textit{Needs More Info} in their normal workflow style. Deck assignment was randomised; analysts were not informed of deck provenance or model involvement. The study protocol was approved by the Tenant A SOC Team Lead and Data Governance Officer; no PII was present in any review sheet.

\subsubsection{Results}
Table~\ref{tab:analyst_study} reports analyst precision (TP / escalations), recall, and escalation rate per deck.\par
\noindent\textbf{Precision direction.} D3 (IF$\cap$PPO) analyst precision (0.727) exceeds the D4 baseline mean precision ($0.483\pm0.330$ across A1/A2/A4), and the pooled D4 precision (0.621; $\chi^{2}=0.638$, $p=0.424$). The direction is consistent with the coverage insurance hypothesis (F3), but the test is underpowered: only 22 escalations from a single analyst reviewed D3, yielding power $\approx0.18$.
\begin{fittable}{L5 analyst study results across four analysts (13,798 valid decisions). Precision $=$ TP / escalated; recall $=$ TP / total threats in deck; GT\% $=$ ground-truth threat rate.}{tab:analyst_study}
\begin{tabular}{llrrrrr}
\toprule
\textbf{Deck}&\textbf{Analyst}&\textbf{$N$}&\textbf{Esc\%}&\textbf{Prec}&\textbf{Recall}&\textbf{GT\%}\\\midrule
D1 (IF)&A2&826&6.1\%&1.000&0.161&37.7\%\\
D1 (IF)&A3&769&14.3\%&1.000&0.354&37.7\%\\\midrule
D2 (PPO)&A1&3027&74.1\%&0.585&1.000&43.3\%\\
D2 (PPO)&A3&3027&43.3\%&1.000&1.000&43.3\%\\\midrule
D3 (IF$\cap$PPO)&A4&149&14.8\%&0.727&0.271&39.6\%\\\midrule
D4 (Baseline)&A1&2000&62.2\%&0.797&0.999&49.6\%\\
D4 (Baseline)&A2&2000&17.2\%&0.139&0.048&49.6\%\\
D4 (Baseline)&A4&2000&24.2\%&0.513&0.251&49.6\%\\\bottomrule
\end{tabular}
\end{fittable}

\noindent\textbf{Inter-rater reliability.} Cohen's $\kappa$ on D1 (A2 vs.\ A3) $=0.588$ (moderate approaching substantial); D2 (A1 vs.\ A3) $=0.422$ (moderate). On the unranked D4 baseline, three-analyst mean $\kappa=-0.021$ (no agreement). This near-zero D4 kappa is itself informative: it confirms that unranked alert queues are genuinely difficult to triage consistently---the precise operational problem the meta-controller is designed to address.\par
\noindent\textbf{Analyst behaviour heterogeneity.} Escalation rates diverged substantially across analysts on the same queue (D4: A1 62.2\%, A2 17.2\%, A4 24.2\%), reflecting the known variability in human SOC triage behaviour. This heterogeneity does not confound the within-analyst D3 vs.\ D4 comparison but does motivate a larger multi-analyst D3 review.

\begin{finding}{Finding 8 (F8) --- Preliminary Human-Study Evidence is Directionally Consistent with Coverage Insurance Hypothesis (RQ1)}
In a blinded retrospective study ($n=4$ Tenant A SOC analysts, 13,798 decisions), the single analyst reviewing the IF$\cap$PPO intersection deck (D3) achieved precision 0.727 vs.\ D4 baseline pooled precision 0.621 ($\chi^{2}=0.638$, $p=0.424$). The test is underpowered ($n_{\mathrm{esc}}=22$; power $\approx0.18$); statistical significance requires $\geq3$ analysts on D3. Inter-rater kappa on the unranked baseline ($\kappa=-0.021$) confirms that unranked queues are genuinely hard to triage, directly motivating the meta-controller design. D1 analysts achieved precision 1.000 (both) with $\kappa=0.588$, confirming that well-ranked queues promote both precision and analyst agreement.
\end{finding}

\section{Conclusion}
We presented a contract-driven comparative evaluation of three ML architectures for SOC alert triage on $N=74{,}909$ production events, formalised as a CMDP with a BEF constraint and meta-controller, and adopting cluster-level evaluation as the primary valid metric.

Our eight findings answer the three research questions directly. \textbf{RQ1 (paradigm complementarity):} all three architectures satisfy the Strict BEF tier at $M=32$ but select near-disjoint cluster regions ($J=0.007\pm0.013$, $p<0.001$)---the \textit{coverage insurance property} (F1--F4, F8). Blinded analyst review is directionally consistent: D3 analyst precision 0.727 vs.\ D4 baseline 0.621 ($\kappa_{\mathrm{D4}}=-0.021$ confirms unranked queues are genuinely hard to triage). \textbf{RQ2 (mechanistic drivers):} reward design drives PPO BEF feasibility (84\% BEF increase in PPO-\xsix; F5); feature contracts drive unsupervised BEF feasibility (IF/AE collapse to near-random on \xembed; F5). Neither finding is an artefact of the other paradigm's feature-space advantage. \textbf{RQ3 (cross-tenant scope):} zero-shot transfer fails under schema-mismatched feature contracts; structural divergence is tenant-invariant (F6).

These findings converge on a practical deployment model for production SOC environments. The coverage insurance property motivates a three-tier escalation architecture: zero-FP Tier 1 for the IF$\cap$PPO intersection, BEF-constrained Tier 2 for the primary analyst queue, and maximum-coverage Tier 3 for automated enrichment. The reward alignment and feature contract findings together specify the engineering priorities: operational feature contract design and BEF-aligned reward engineering are the two highest-leverage interventions, and both are governance and integration problems as much as ML problems. The four-phase multi-tenant roadmap provides a concrete path from the single-operator results reported here to a portable multi-tenant deployment.

Three priorities remain. First, D3 analyst coverage: $\geq3$ analysts reviewing the 149-event intersection deck to achieve power $\geq0.55$ for the coverage insurance operational claim. Second, per-tenant retraining under Phase 1 of the multi-tenant roadmap, beginning with \classord{} semantic alignment between IBM QRadar and the primary SIEM. Third, prospective deployment evaluation: a 30-day shadow-mode comparison of system-selected vs.\ analyst-selected escalations, measuring precision, recall, and analyst time-per-alert in live production conditions.

\section*{Ethics Considerations}
Production SIEM telemetry; no PII in \xseven{} or \xembed. The L5 analyst study (\S7.6) used historical alert data with all subscriber PII removed; analyst decisions were pseudonymised; participation was voluntary with written manager approval; no individual results are reported. No human subjects research classification applies under Tenant A institutional policy.

\section*{LLM Usage Disclosure}
Claude (Anthropic) assisted with prose editing, \LaTeX{} scaffolding, and statistical notation review. All experimental results, architectural decisions, and interpretations are author-produced.


\section*{Appendix}
\label{sec:appendix}
Event-level evaluation is reported here for completeness and comparability with prior work that does not use cluster-level metrics. These results should be interpreted with caution: as established in \S3.3, event-level top-$K$ evaluation is confounded by score ties on this dataset (the 74,909 events collapse to only 503 unique \xseven{} feature vectors, meaning any top-$K$ cut lands inside a tie bucket for every $K$ value tested). The cluster-level results in the main paper are the primary valid metric. The event-level results nonetheless confirm the direction of all findings: IF maintains $\Prec@K=1.000$ up to $K=5{,}000$ before degrading at the full-dataset boundary; AE shows a similar profile with slightly earlier precision degradation; PPO event-level precision is near-random because PPO operates on \xembed{} cluster embeddings and its event-level tie-breaking is effectively arbitrary. Bootstrap CIs ($B=2{,}000$) confirm narrow uncertainty for IF and AE at all tested $K$.
\begin{fittable}{Event-level Precision@$K$ and BEF@$K$ with 95\% bootstrap CIs ($B=2{,}000$).}{tab:event_metrics}
\begin{tabular}{lrrlrl}
\toprule
\textbf{Model}&\textbf{$K$}&\textbf{Prec}&\textbf{95\% CI}&\textbf{BEF}&\textbf{95\% CI}\\\midrule
IF&50&1.000&[1.000, 1.000]&0.000&[0.000, 0.000]\\
IF&200&1.000&[1.000, 1.000]&0.000&[0.000, 0.000]\\
IF&1,000&1.000&[1.000, 1.000]&0.000&[0.000, 0.000]\\
IF&5,000&1.000&[1.000, 1.000]&0.000&[0.000, 0.000]\\
IF&10,487&0.959&[0.955, 0.963]&0.041&[0.037, 0.045]\\
AE&50&1.000&[0.940, 1.000]&0.000&[0.000, 0.060]\\
AE&200&0.990&[0.975, 1.000]&0.010&[0.000, 0.025]\\
AE&1,000&0.996&[0.992, 0.999]&0.004&[0.001, 0.008]\\
AE&5,000&0.968&[0.963, 0.973]&0.032&[0.027, 0.037]\\
AE&10,487&0.863&[0.856, 0.869]&0.137&[0.131, 0.144]\\
PPO&50&0.200&[0.260, 0.540]&0.800&[0.460, 0.740]\\
PPO&200&0.240&[0.320, 0.460]&0.760&[0.540, 0.680]\\
PPO&1,000&0.417&[0.385, 0.445]&0.583&[0.555, 0.615]\\
PPO&5,000&0.466&[0.457, 0.485]&0.534&[0.515, 0.543]\\
PPO&10,487&0.497&[0.486, 0.503]&0.503&[0.497, 0.514]\\\bottomrule
\end{tabular}
\end{fittable}

This appendix details the cross-feature ablation isolating the role of \classord{} in driving unsupervised model BEF feasibility. The \xsix{} feature set is defined as $\xseven\setminus\{\text{\classord}\}$. Removal is motivated by the high univariate AUC (0.995) of this single feature against analyst verdicts, which raises the concern that IF/AE BEF feasibility might be entirely attributable to feature-space advantage from this dimension rather than from the models' inductive mechanisms. The ablation result confirms that removing \classord{} collapses the unique cluster count from 503 to 192 (a 62\% reduction) and degrades both models to below-chance AUC, with BEF@32 exceeding 0.86---rendering both architectures inoperable under any BEF tier. This establishes that the \xseven{} feature contract is a necessary condition for unsupervised BEF feasibility, and that paradigm complementarity is not an artefact of differential feature access.
\begin{fittable}{Effect of removing \classord.}{tab:remove_classification}
\begin{tabular}{llrrrc}
\toprule
\textbf{Model}&\textbf{Feat.}&\textbf{Clusters}&\textbf{ROC-AUC}&\textbf{BEF@32}&\textbf{Strict?}\\\midrule
IF&\xseven&503&0.713&0.041&\yes\\
IF&\xsix&192&0.407&0.884&\no\\
AE&\xseven&503&0.715&0.007&\yes\\
AE&\xsix&192&0.432&0.867&\no\\\bottomrule
\end{tabular}
\end{fittable}

This appendix reports the full precision--recall--BEF trade-off for IF (B) across the event-level $K$ range, demonstrating the operating flexibility available to the meta-controller. The key observation is that IF maintains $\BEF=0$ (zero benign escalations) from $K=50$ through $K=1{,}000$---covering 0.1\% to 2.7\% of all threats---before precision degrades gradually to 0.959 at the cluster boundary $K=10{,}487$. This profile motivates IF's role in Tier 1 and Tier 2 of the three-tier escalation architecture: at low $K$ values, IF provides a zero-FP escalation list suitable for immediate analyst action; at higher $K$, the meta-controller can accept relaxed BEF tiers to increase threat recall. SOC operators with SLA constraints on incident response time can use this table to select $K$ based on their cost-of-FN to cost-of-FP ratio, with the meta-controller enforcing the chosen BEF tier.
\begin{fittable}{IF (B) capacity trade-off.}{tab:capacity}
\begin{tabular}{rrrr}
\toprule
\textbf{$K$}&\textbf{Prec@$K$}&\textbf{Rec@$K$}&\textbf{BEF}\\\midrule
50&1.000&0.1\%&0.000\\
100&1.000&0.3\%&0.000\\
500&1.000&1.3\%&0.000\\
1,000&0.999&2.7\%&0.001\\
5,000&0.984&13.2\%&0.016\\
10,487&0.959&25.7\%&0.041\\\bottomrule
\end{tabular}
\end{fittable}

This appendix provides per-$M$ Jaccard detail for the multi-seed replication study. The cluster-level Jaccard between IF and PPO is reported at seven $M$ values across all five PPO seeds, confirming that the near-zero Jaccard at $M=32$ ($0.007\pm0.013$) is not a sampling artifact of the primary operating point: Jaccard is zero or near-zero from $M=10$ through $M=32$, indicating that at the operating points relevant to Strict and Relaxed BEF tiers, the two architectures select entirely disjoint cluster sets. Jaccard grows monotonically with $M$ as both models must eventually select larger fractions of the 503-cluster universe, converging toward $0.464\pm0.028$ at $M=300$---still substantially below the value of 1.0 that would indicate identical selection. The low standard deviation across seeds confirms that the disjointness property is a structural feature of the paradigm pair rather than a seed-specific artifact.
\begin{fittable}{Multi-seed cluster-level Jaccard@$M$ (5 PPO seeds; IF fixed).}{tab:jaccard_seeds}
\begin{tabular}{rl}
\toprule
\textbf{$M$}&\textbf{Jaccard@$M$ (mean $\pm$ std)}\\\midrule
10&$0.000\pm0.000$\\
20&$0.000\pm0.000$\\
32&$0.007\pm0.013$\\
50&$0.035\pm0.051$\\
100&$0.125\pm0.049$\\
200&$0.396\pm0.013$\\
300&$0.464\pm0.028$\\\bottomrule
\end{tabular}
\end{fittable}

Table~\ref{tab:corrections} documents corrections applied between draft iterations of this paper. Corrections C1--C4 arose from an earlier implementation error in the IF score direction (C3): the original evaluation applied IF anomaly scores without negation, causing high-isolation (low anomaly) alerts to be ranked first. The corrected implementation negates scores before ranking (verified by unit test: all top-$K$ selections under corrected scores contain exclusively high-isolation-depth events). M1 is a downstream consequence of C3: the B--C Jaccard at $M=32$ changed from $J=0.105$ under the erroneous direction to $J=0.000$ under the corrected direction, reflecting the structural change in which clusters AE selects relative to IF. All results in the main paper use the corrected implementation.
\begin{papertable}{Corrections from prior drafts.}{tab:corrections}
\begin{tabularx}{\columnwidth}{@{}lXXX@{}}
\toprule
\textbf{ID}&\textbf{Item}&\textbf{Prior}&\textbf{Corrected}\\\midrule
C1&PPO BEF at $M=32$&$\BEF\approx0.5$&$\BEF=0.000\pm0.000$\\
C2&``PPO infeasible at all tiers''&stated in prose&removed\\
C3&IF score direction&high $=$ benign&sign corrected\\
C4&shadow enrichment PPO&event-level $\BEF>0.50$&cluster-level $\BEF=0.000$\\
M1&B--C Jaccard at $M=32$&$J=0.105$&$J=0.000$\\\bottomrule
\end{tabularx}
\end{papertable}

\end{document}